 \documentclass[aps,pra,twocolumn,superscriptaddress,floatfix,showpacs]{revtex4-1}
 
\usepackage{bm,bbm,amsmath, amssymb,amsbsy,amsthm,graphicx,subfigure,color,txfonts,multirow}
\usepackage{amsmath}
\usepackage{amsfonts}
\usepackage{array}

\usepackage[utf8x]{inputenc}
\usepackage[greek,english]{babel}
\usepackage{teubner}

\usepackage[framemethod=tikz]{mdframed}
\usepackage[colorlinks]{hyperref}

\hypersetup{
	bookmarksnumbered,
	pdfstartview={FitH},
	citecolor={blue},
	linkcolor={red},
	urlcolor={black},
	pdfpagemode={UseOutlines}
}
 
\newcommand{\kebra}[2]{\vert{#1}\rangle\langle{#2}\vert}
 \newcommand{\tr}[1]{\text{Tr}}
\newcommand{\ket}[1]{|#1\rangle}
\newcommand{\bra}[1]{\langle#1|}

\begin{document}
\title{Detecting metrologically useful asymmetry and entanglement  by a few local measurements}

\author{Chao Zhang}
\affiliation{Key Laboratory of Quantum Information, University of Science and Technology of China, CAS, Hefei, 230026, China}
\affiliation{Synergetic Innovation Center of Quantum Information and Quantum Physics, University of Science and Technology of China, Hefei, 230026, P.R. China}
\author{Benjamin Yadin}\affiliation{Department of Atomic and Laser Physics, University of Oxford, Parks Road, Oxford OX1 3PU, United Kingdom}
\author{Zhi-Bo Hou}
\affiliation{Key Laboratory of Quantum Information, University of Science and Technology of China, CAS, Hefei, 230026, China}
\affiliation{Synergetic Innovation Center of Quantum Information and Quantum Physics, University of Science and Technology of China, Hefei, 230026, P.R. China}
\author{Huan Cao}
\affiliation{Key Laboratory of Quantum Information, University of Science and Technology of China, CAS, Hefei, 230026, China}
\affiliation{Synergetic Innovation Center of Quantum Information and Quantum Physics, University of Science and Technology of China, Hefei, 230026, P.R. China}
\author{Bi-Heng Liu}
\affiliation{Key Laboratory of Quantum Information, University of Science and Technology of China, CAS, Hefei, 230026, China}
\affiliation{Synergetic Innovation Center of Quantum Information and Quantum Physics, University of Science and Technology of China, Hefei, 230026, P.R. China}
\author{Yun-Feng Huang}
\email{hyf@ustc.edu.cn}
\affiliation{Key Laboratory of Quantum Information, University of Science and Technology of China, CAS, Hefei, 230026, China}
\affiliation{Synergetic Innovation Center of Quantum Information and Quantum Physics, University of Science and Technology of China, Hefei, 230026, P.R. China}
\author{Reevu Maity}\affiliation{Department of Atomic and Laser Physics, University of Oxford, Parks Road, Oxford OX1 3PU, United Kingdom}
\author{Vlatko Vedral}\affiliation{Department of Atomic and Laser Physics, University of Oxford, Parks Road, Oxford OX1 3PU, United Kingdom}
\affiliation{Centre for Quantum Technologies, National University of Singapore, 117543  Singapore}
\author{Chuan-Feng Li}
\email{cfli@ustc.edu.cn}
\affiliation{Key Laboratory of Quantum Information, University of Science and Technology of China, CAS, Hefei, 230026, China}
\affiliation{Synergetic Innovation Center of Quantum Information and Quantum Physics, University of Science and Technology of China, Hefei, 230026, P.R. China}
\author{Guang-Can Guo}
\affiliation{Key Laboratory of Quantum Information, University of Science and Technology of China, CAS, Hefei, 230026, China}
\affiliation{Synergetic Innovation Center of Quantum Information and Quantum Physics, University of Science and Technology of China, Hefei, 230026, P.R. China}
 \author{Davide Girolami}
\email{davegirolami@gmail.com}
\affiliation{Department of Atomic and Laser Physics, University of Oxford, Parks Road, Oxford OX1 3PU, United Kingdom}

\begin{abstract}
Important properties of a quantum system are not directly measurable, but they can be disclosed by how fast the system changes under controlled perturbations. In particular, asymmetry and entanglement can be verified by reconstructing the state of a quantum system. Yet, this usually requires experimental and computational resources which increase exponentially with the system size. Here we show how to detect metrologically useful asymmetry and entanglement by a limited number of measurements. This is achieved by studying how they affect the speed of evolution of a system under a unitary transformation. We show that the speed of multiqubit systems can be evaluated by measuring a set of local observables, providing exponential advantage with respect to state tomography. Indeed, the presented method requires neither the knowledge of the state and the parameter-encoding Hamiltonian nor global measurements performed on all the constituent subsystems. We implement the detection scheme in an all-optical experiment.
\end{abstract}


\date{\today}

\maketitle
\section{Introduction} Quantum coherence and entanglement can generate  non-classical speed-up in information processing \cite{nielsen}. Yet, their experimental verification  is challenging.  Being not directly observable, their detection usually implies reconstructing the full state of the system, which requires a number of measurements growing exponentially with the system size \cite{bookest}. Also, verifying their presence is necessary, but not sufficient to guarantee a computational advantage.  \\
Here we show how to detect {\it useful} coherence and entanglement  in systems of arbitrary dimension by  a limited sequence of measurements.  We propose an experimentally friendly measure of the speed of  a quantum system, i.e. how fast its state changes under a generic channel, which for $n$-qubit systems is a function of a linearly scaling $(O(n))$ number of observables. 
The speed of  a quantum system  determines its computational power  \cite{metrorev,levitin,spekkens0,lloyd}. Quantum speed limits of open systems also provides information about the environment structure \cite{taddei,delcampo,speedger}, helping develop efficient control strategies \cite{toth,caneva,deffner,control}, and investigate phase transitions  of condensed matter systems   \cite{critic2,zanardi}.
We prove a quantitative link between our speed measure, when undertaking a unitary dynamics, and metrological quantum resources. In Section \ref{two}, we relate speed to asymmetry, i.e. the coherence with respect to an Hamiltonian eigenbasis. Asymmetry underpins the usefulness of a probe  to phase estimation and reference frame alignment  \cite{vaccaro,asymrev,spekkens2,spekkens}.  Moreover,  a superlinear scaling of the speed of multipartite systems certifies an advantage in metrology  powered by entanglement, as discussed in Section \ref{three}.  We show how to detect asymmetry and entanglement by comparing the speed of two copies of a system, while performing a phase encoding dynamics on only one copy. An important advantage of the method is that a priori knowledge of the input state and the Hamiltonian is not required. We demonstrate the scheme in an all-optical experiment, described in Section \ref{four}.  An asymmetry lower bound and an  entanglement witness are extracted from the speed of  a two-qubit system in dynamics generated by additive spin Hamiltonians, without brute force state reconstruction. In Section \ref{five}, we provide for the interested reader a brief review of information geometry concepts and the complete proofs of the theoretical results. We draw our conclusions in Section \ref{six}.  \\

 \section{Relating asymmetry to observables}\label{two}
  The sensitivity of  a quantum system to a quantum operation  described by a parametrized channel  $\Phi_t$ \cite{nielsen}, where $t$ is the time, is determined by how fast its state $\rho_t:=\Phi_t(\rho_0)$ evolves. We quantify the system speed over an interval $0\leq t\leq \tau$  by the average rate of change of the state, which is given by mean values of quantum operators $\langle\cdot\rangle_{\rho_t}=\text{Tr}(\cdot\rho_t)$:
\begin{eqnarray}\label{eq1}
s_{\tau}(\rho_t):=\frac{||\rho_\tau-\rho_0||_2}{\tau}= \frac{(\langle\rho_\tau\rangle_{\rho_\tau}+\langle\rho_0\rangle_{\rho_0}-2\langle\rho_\tau\rangle_{\rho_0})^{1/2}}{\tau}, 
\end{eqnarray}
where the Euclidean distance is employed. 
  Measuring the swap operator on two system copies is sufficient to quantify state overlaps,  $\langle\sigma\rangle_\rho=\langle V \rangle_{\rho\otimes\sigma},  V(\ket{\phi_1}\otimes\ket{\phi_2})=\ket{\phi_2}\otimes\ket{\phi_1}, \forall \ket{\phi_{1,2}}$.  
The global swap is the product of local swaps,  $V_{S}=\otimes_{i=1}^nV_{S_i}$. Then,  for $n$-qubit  systems  $S\equiv\{S_i\}, i=1,\ldots,n$, a state overlap $\langle\sigma_S\rangle_{\rho_{S}}$ is obtained by evaluating  $O(n)$ observables, one for each pair of  subsystem $S_i$ copies \cite{alves,jeong,greiner}.  Each local swap can be recast in terms of projections on the Bell singlet $V_{S_i}=I_{d^2}-2\Pi_{S_i}^{\psi^-}, \Pi_{S_i}^{\psi^-}=\ket{\psi^-}\bra{\psi^-}_{S_i},\ket{\psi^-}=1/\sqrt2(\ket{01}-\ket{10})$, a standard routine of quantum information processing, e.g. in bosonic lattices. Bell state projections are implemented by $n$  beam splitters  interfering each pair of $S_i$ copies,  and coincidence detection on the correlated pairs. Hence, the speed of  an $n$-qubit system  is evaluated by networks whose size  scales linearly with the number of subsystems, employing $O(n)$ two-qubit gates and  detectors. Note that  tomography demands to prepare $O(2^{2n})$  system copies and perform a measurement on each of them \cite{bookest}.  It is also possible to extract the swap value  by single qubit interferometry \cite{brun,ekert,filip}. The two copies of the system are correlated with an ancillary qubit by a controlled-swap gate.
  The mean value of the swap is then encoded in the ancilla polarisation. Yet, the implementation of a controlled-swap gate is currently a serious challenge \cite{fredkin}.      \\
 


Crucial properties of quantum systems can be determined by measuring the speed defined in Eq.~(\ref{eq1}), without  further data. 
 Performing a quantum computation $U_t\rho U_t^{\dagger}, U_t=e^{-i H t},$ relies on the coherence in the Hamiltonian $H$ eigenbasis, a property called ($U(1)-$) asymmetry \cite{spekkens2,spekkens,asymrev,vaccaro}. In fact, incoherent states in such a basis do not evolve. Asymmetry is operationally defined as the system ability to break a symmetry generated by the Hamiltonian. 
Asymmetry measures are defined as non-increasing functions in symmetry-preserving dynamics, which are modelled by transformations $\Phi$ commuting with the Hamiltonian evolution, $[\Phi,U_t]=0$.\\
  Experimentally measuring coherence, and in particular asymmetry, is hard \cite{cohrev,me}. One cannot discriminate with certainty coherent states from incoherent mixtures, without full state reconstruction. We show how to evaluate the asymmetry of a system by its speed (full details and proofs in Section~\ref{five}).  To quantify the sensitivity of a probe state $\rho=\sum_i\lambda_i\ket i\bra i$ to the unitary transformation $U_{t}$, we employ  the symmetric logarithmic derivative quantum Fisher information (SLDF), a widely employed quantity in quantum metrology and quantum information \cite{helstrom}:
\begin{eqnarray}
 {\cal I}_F(\rho, H)=1/2\sum_{i,j} \frac{(\lambda_i-\lambda_j)^2}{\lambda_i+\lambda_j} |\langle i|H| j\rangle|^2. 
\end{eqnarray}
Note that the SLDF is one of the many quantum extensions of the classical Fisher information \cite{petzmono}. Indeed, the SLDF is an ensemble asymmetry monotone, i.e. an asymmetry measure, being contractive  on average under commuting operations \cite{ben}:
\begin{eqnarray}
  {\cal I}_F(\rho,H)&\geq&\sum_{\mu}p_{\mu} {\cal I}_F(\Phi_{\mu}(\rho),H),\\
 \forall \{p_{\mu},\Phi_{\mu}\} &:&\sum_{\mu}p_{\mu}=1, [\Phi_{\mu},U_t]=0. \nonumber 
\end{eqnarray}
We observe that this implies that every quantum Fisher information is an asymmetry ensemble monotone, see Section \ref{five}.

  Reconstructing both  state and Hamiltonian is  required to compute the SLDF. Yet, few algebra steps show  that it is lower bounded by the squared speed over an interval $\tau$ of the evolution $U_t \rho U_t^{\dagger}$:
\begin{eqnarray}\label{eq4}
 {\cal S}_{\tau}(\rho,H):&=&s_{\tau}(\rho)^2/2= \frac{\langle\rho\rangle_{\rho}  -\langle U_{\tau}\rho U_{\tau}^{\dagger}\rangle_{\rho}}{\tau^2}, \\ 
 {\cal S}_{\tau}(\rho,H)&\leq& {\cal I}_F(\rho,H), \ \ \forall \rho, \tau, H, \nonumber
\end{eqnarray}
  where we drop the time label, as the speed  is constant. It is then possible to  bound asymmetry with respect to an arbitrary Hamiltonian by evaluating the purity $\langle\rho \rangle_{\rho}$ and the overlap $\langle  U_{\tau}\rho U_{\tau}^{\dagger}\rangle_{\rho}$. A non-vanishing  speed reliably witnesses asymmetry, $s_{\tau}(\rho)>0 \iff {\cal I}_F(\rho,H)>0, \forall \tau$. 
  The Hamiltonian variance is an upper bound to asymmetry, up to a constant, ${\cal I}_F(\rho,H)\leq{\cal V}(\rho,H)=\langle H^2\rangle_{\rho}-\langle H\rangle_{\rho}^2, \forall  \rho, H$. Yet, the variance is generally not a reliable indicator of asymmetry, as it is arbitrarily large for incoherent mixed states. The chain of inequalities is saturated for pure states, in the zero time limit, $\lim\limits_{\tau\rightarrow 0}{\cal S}_{\tau}(\rho_{\psi},H)= {\cal I}_F(\rho_\psi,H)={\cal V}(\rho_\psi,H), \rho_{\psi}=\ket{\psi}\bra{\psi}$. \\ 
  In fact, the quantum Fisher informations quantify the instantaneous  response to a perturbation \cite{toth,petzmono}. \\
 
 \section{Witnessing metrologically useful entanglement}\label{three}
 We extend the analysis to multipartite systems, proving that  non-linear speed scaling witnesses useful  entanglement.  Consider a phase estimation protocol, a  building block of quantum computation and metrology schemes \cite{nielsen,toth,metrorev}. A phase shift $U_{\tau,i}=e^{-i h_i\tau}$ is applied in parallel to each site of an $n$-qubit probe. The  generator is an additive Hamiltonian $H_n=\sum_{i=1}^n h_i, h_i=I_{1,\ldots,i-1}\otimes\sigma_i\otimes I_{i+1,\ldots,n},$ where $\sigma$ is an arbitrary spin-1/2 operator.
   The goal is to estimate the parameter $\tau$ by an estimator $\tau_{\text{est}}$, extracted from measurements on the perturbed system.  The quantum Cram\'{e}r-Rao bound establishes that asymmetry, measured by the SLDF, bounds the estimation precision, expressed via the estimator variance, ${\cal V}(\rho,\tau_{\text{est}})\geq  (\nu {\cal I}_{F}(\rho, H_n))^{-1}, \forall \rho, H_n,$ where $\nu$ is the number of trials, and  the estimation is assumed unbiased, $\langle\tau_{\text{est}}\rangle_{\rho}=\tau$. Separable states achieve at best ${\cal I}_F(\rho, H_n)= O(n)$, while entanglement asymptotically enables up to a quadratic improvement, $ {\cal I}_F(\rho, H_n)=O(n^2), n\rightarrow \infty$.  Specifically, with the adopted normalization, the relation ${\cal I}_F(\rho, H_n)> n/4$, i.e. super-linear asymmetry with respect to an additive observable,  witnesses  entanglement \cite{pezze}.   Given Eq.~(\ref{eq4}),  entanglement-enhanced precision in estimating a phase shift $\tau$ is verified if
\begin{equation}\label{eq5}
{\cal S}_{\tau}(\rho, H_n)> n/4.
\end{equation} 
 \begin{figure}[t]
  \centering
 \includegraphics[height=6.5cm,width=8.5cm]{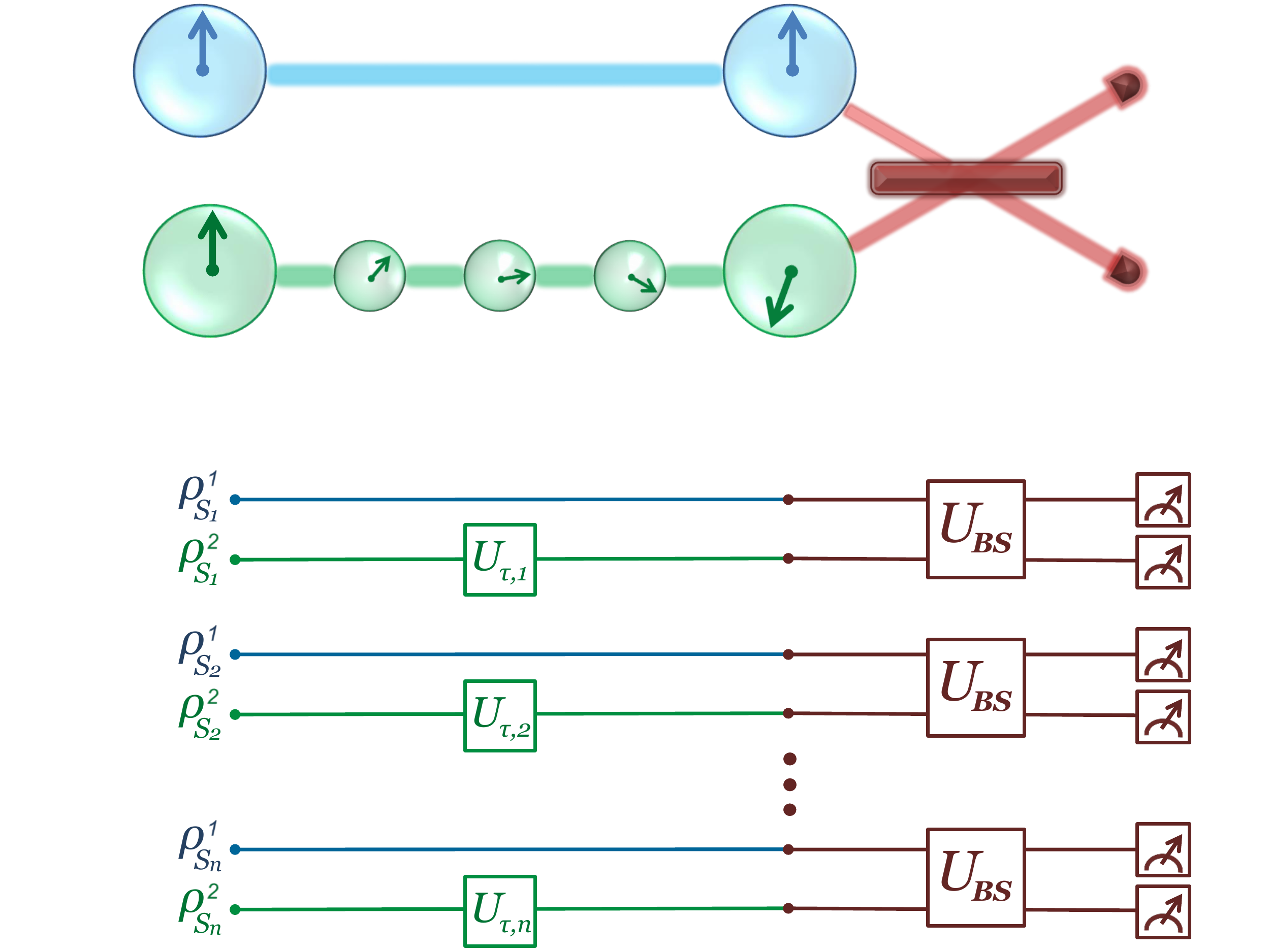}
\caption{\label{fig1}{\bf Overlap detection.} The network evaluates the overlap $\langle e^{-i H_n \tau}\rho^{2}_{S}e^{i H_n \tau}\rangle_{\rho^1_S}, H_n=\sum_{i=1}^n h_i,$ in an $n$-qubit system $S\equiv\{S_i\}$.  Each pair of subsystem $S_i$ copies, in the state $\rho^{1}_{S_{i}}\otimes\rho^{2}_{S_{i}}$, enters a two-arm channel (blue and green).  The unitaries $U_{\tau,i}=e^{-i h_i \tau}$ are applied to the second copy of each pair. Leaving both copies unperturbed, the network measures the state purity. The measurement apparatus (red) interferes each pair of subsystem copies by $O(n)$ beam splitter gates $U_{BS}$ \cite{jeong}. The overlap, and therefore the speed function in Eq.~\ref{eq4}, is extracted by $O(n)$ local detections.    
} 
\end{figure}
 The overlap detection network  for $n$-qubit systems and additive Hamiltonians is depicted in Fig.~\ref{fig1}. 
Evaluating the SLDF is an appealing strategy to verify an advantage given by  entanglement, rather than just detecting quantum correlations \cite{tomonec,multi1,multi2,multi3,multi4,greiner,entropy}.  The SLDF of thermal states can be extracted by measuring the system  dynamic susceptibility \cite{zoller}, while lower bounds are obtained by two-time detections of a global observable \cite{scienceexp,frowis}.  Also,  collective observables can witness entanglement in highly symmetric states \cite{apel}. 
  \begin{figure}[t]
\centering
\includegraphics[width=8cm,height=7cm]{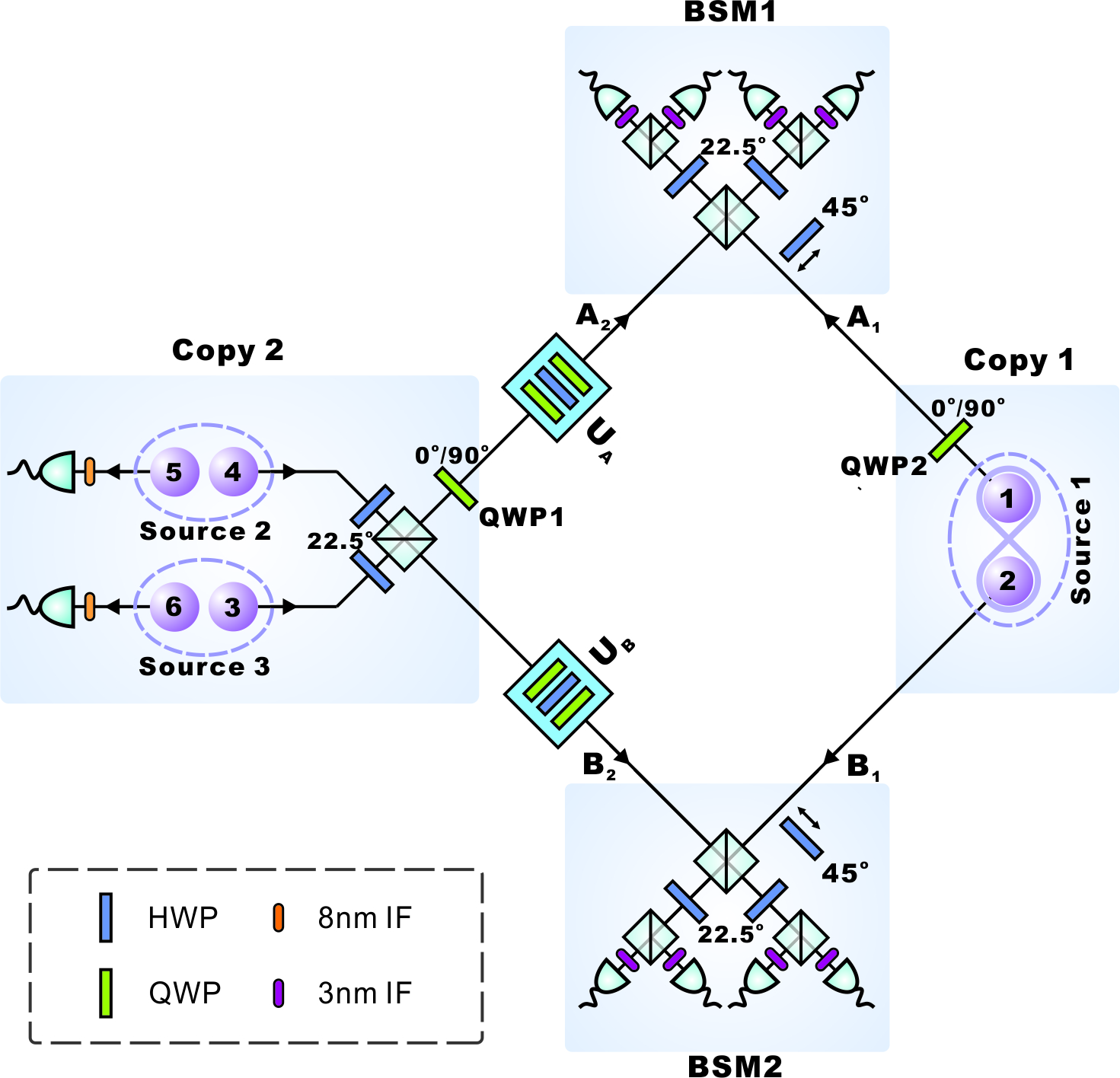}
\caption{\label{setup} {\bf Experimental set-up}. We prepare two copies of a Bell state $\ket{\phi^+}$ by a laser-emitted ultraviolet pulse split into three beams pumping  SPDC sources.   The scheme guarantees that both copies are emitted by   different sources. Conversely, in a two source setting, the fourfold coincidences in the BSMs could be generated by two photon pairs emitted from a single source, invalidating the experiment.   The four terms of the mixture are obtained by rotating QWP1,2. Purity and overlap measurements are implemented via BSM schemes.  A multi-channel unit counts the sixfold coincidences (one detector fire in each output mode).}
\end{figure}
Our proposal has two peculiar advantages. First, it is applicable to any probe state $\rho$ without a priori information and assumptions, e.g.  invariance under permutation of the subsystems. Second, only local pairwise interactions and detections are needed. This means that distant laboratories  can verify quantum speed-up due to entanglement  in a shared system $S$  by local operations and classical communication  \cite{nielsen}, providing each laboratory  with two copies of a subsystem $S_i$. Note that quadratic speed  scaling certifies the probe optimization, ${\cal S}_{\tau}(\rho, H_n)=O(n^2) \Rightarrow {\cal I}_F(\rho, H_n)=O(n^2)$.  \\

\section{Experimental asymmetry and entanglement detection}\label{four}
\subsection{Implementation}
 We experimentally extract a lower bound to  metrologically useful   asymmetry and entanglement of a two-qubit system $AB$  in an optical set-up, by measuring its speed during a unitary evolution. While employing state tomography would require fifteen measurements, we verify that the proposed protocol needs six. 
The system is prepared in a mixture of Bell states, $\rho_{p,AB}=p \ket{\phi^+}\bra{\phi^+}+(1-p) \ket{\phi^-}\bra{\phi^-},  \ket{\phi^{\pm}}=1/\sqrt2 (\ket{00}\pm \ket{11}), p\in [0,1].$   We implement transformations generated by the Hamiltonians $H_2=\sum_{i=A,B} h_i, h=  \sigma_{x,y,z}$, where $\sigma_{x,y,z}$ are the spin-1/2 Pauli matrices, for  equally stepped values of the mixing parameter, $p=0,0.1,0.2,\ldots,0.9,1,$ over an interval $\tau=\pi/6$.  The squared speed function $S_{\pi/6}(\rho_p, H_2)$ is evaluated  from purity and overlap measurements. \\
Each run of the experiment implements the  scheme in Fig.~\ref{setup}. We prepare two copies (Copy 1,2) of a maximally entangled two-qubit state $\ket{\phi^+}=1/\sqrt{2}({\bf HH+VV})$, where ${\bf H,V}$ label horizontal and vertical photon polarizations, from three  spontaneous parametric down-conversion  sources (SPDC Source 1,2,3). They are generated by ultrafast 90 mW pump pulses from a mode-locked Ti:Sapphire laser, with a central wavelength of 780 nm, a pulse duration of 140 fs, and a repetition rate of 76 MHz.  
Copy 1 (photons 1,2) is obtained from Source 1, by employing a sandwich-like Beta-barium Borate (BBO) crystal \cite{source}. Copy 2  is prepared from Source 2,3.  Two photon pairs (photons 3-6) are generated via single BBO crystals (beamlike type-II phase matching). By detecting photons 5,6, a product state encoded in photons 3,4 is triggered. Photons 3-4 polarisations are rotated via half-wave plates (HWPs). 
They are then interfered by a polarizing beam splitter (PBS) for parity check measurements. 
We then simulate the preparation of the state $\rho_{p}^1\otimes \rho^2_{p}  = p^2\Pi^{\phi^+\phi^+}_{12}+p(1-p)(\Pi^{\phi^+\phi^-}_{12}+\Pi^{\phi^-\phi^+}_{12})+(1-p)^2\Pi^{\phi^-\phi^-}_{12}, \Pi^{\phi^{\pm}\phi^{\pm}}_{12}=|\phi^{\pm}\rangle\langle\phi^{\pm}|_{A_1B_1}\otimes|\phi^{\pm}\rangle\langle\phi^{\pm}|_{A_2B_2}$.   Classical mixing is obtained by applying quarter-wave plates (QWP1,QWP2)  to each system copy.  A $90^\circ$ rotated QWP swaps the Bell states, $|\phi^{\pm}\rangle\rightarrow|\phi^{\mp}\rangle$, generating a $\pi$ phase shift between ${\bf H, V}$ polarisations.  The four terms of the mixture are obtained in separate runs by engineering the rotation sequences  $(\text{QWP1,QWP2})=\{(0^\circ,0^\circ),(0^\circ,90^\circ),(90^\circ,0^\circ),(90^\circ,90^\circ)\},$  with a duration proportional to $\{p^2,p(1-p),p(1-p),(1-p)^2\},$ respectively. The collected data from the four cases are then identical to the ones obtained from direct preparation of the mixture. \\
We quantify the speed by measuring the purity $\langle V_{12}\rangle_{\rho_{p}^1\otimes\rho_{p}^2}$ and the overlap $\langle V_{12}\rangle_{\rho^1_{p}\otimes U_{\pi/6}\rho_{p}^2 U^{\dagger}_{\pi/6}}$.   The unitary gate $U_{\pi/6}=U_{\pi/6,A_2}\otimes U_{\pi/6,B_2}, U_{\pi/6,A_2(B_2)}=e^{-i h_{A_2(B_2)} \pi/6},$  is applied to the second system copy by a sequence of one HWP sandwiched by two QWPs. The sequences of gates implementing each Hamiltonian are obtained as follows. Single qubit unitary gates implement SU(2) group transformations. We parametrize the rotations  by the Euler angles  $(\xi, \eta, \zeta)$:
\begin{equation}
u(\xi, \eta, \zeta):=\exp\left(-i\frac{1}{2}\xi\sigma_y\right)\exp\left(-i\frac{1}{2}\eta \sigma_x\right)\exp\left(-i\frac{1}{2}\zeta\sigma_y\right), 
\end{equation}
where $\sigma_{x,y,z}$ are the  Pauli matrices. One can engineer arbitrary single qubit gates by a $\theta$-rotated HWP implementing the transformation $H_\theta$,  sandwiched by two rotated QWPs (transformations $Q_\theta$):
\begin{equation}
u(\xi, \eta, \zeta)=Q_{\theta_3}H_{\theta_2}Q_{\theta_1}, 
\end{equation}
 where $\theta_{1,2,3}$ are the rotation angles to apply to each plate \cite{unitary}. In particular, any unitary transformation is prepared by a gate sequence of the form
\begin{align}
 \theta_1&=\pi/4-\zeta/2\ mod\  \pi,
\nonumber\\
 \theta_2&=-\pi/4+(\xi+\eta-\zeta)/4\ mod\ 
 \pi, \nonumber\\
\theta_3&=\pi/4+\xi/2\ mod\ \pi. 
\end{align}
 The phase shift angles characterising the Hamiltonian evolutions studied in our experiment are shown in Table I.

   \begin{table}
\centering
\begin{tabular}{|c|c|c|c|c|} \hline
Angles & I & $U_X$ & $U_Y$ & $U_Z$ \\\hline
$\theta_1$ & $\frac{\pi}{4}$ & $\frac{\pi}{2}$ & $\frac{\pi}{4}$ & $\frac{\pi}{4}$ \\\hline
$\theta_2$ & $\frac{\pi}{4}$ & $-\frac{\pi}{24}$ & $\frac{5\pi}{24}$ & $\frac{5\pi}{24}$ \\\hline
$\theta_3$ & $\frac{\pi}{4}$ & $\frac{\pi}{2}$ & $\frac{\pi}{6}$ & $\frac{\pi}{4}$ \\\hline
\end{tabular} 
\caption{{\bf Angles of the wave plates implementing the unitary gates.}}
\end{table}

 \begin{table}[t] 
{\footnotesize
\begin{tabular}{|c|c|c|c|} 
\hline
 $h $&$\sigma_x$& $\sigma_y$  &$ \sigma_z$\\
\hline
 ${\cal I}_F(\rho_p, H_2)$& $p$   &    $(1-p)$
&  $(1-2p)^2$    \\
\hline
${\cal S}_{\tau}(\rho_p, H_2)$ &$ (p\sin{\tau}/4\tau)^2$& $((1-p)\sin{\tau}/4\tau)^2 $
 &  $((1-2p)\sin{\tau}/4\tau)^2$   \\
 \hline
${\cal I}_F(\rho_p, H_2)>0.5$ &$p>0.5$& $p<0.5$
 & $p<0.147, p>0.853$  \\
 \hline
  $ {\cal S}_{\pi/6}(\rho_p, H_2)>0.5$&  $p>0.741$  & $p<0.259$  &    $p<0.129, p>0.870$   \\
\hline
\end{tabular} }
\caption{\label{table}{\bf Theoretical values.} The Table reports the theoretical values of the SLDF, which is the smallest quantum Fisher information (multiplying it by a constant turns it into the biggest one, see Section \ref{five}), the lower bound  ${\cal S}_\tau(\rho_p,H_2)$ (Eq.~(\ref{eq4})), and the related entanglement witness conditions (Eq.~(\ref{eq5})), for the Bell state mixture $\rho_p$ and the spin Hamiltonians $H_2$.}
\end{table}
   
\begin{figure*}[th!]
\subfigure[]{
\includegraphics[height=3.7cm,width=5.55cm]{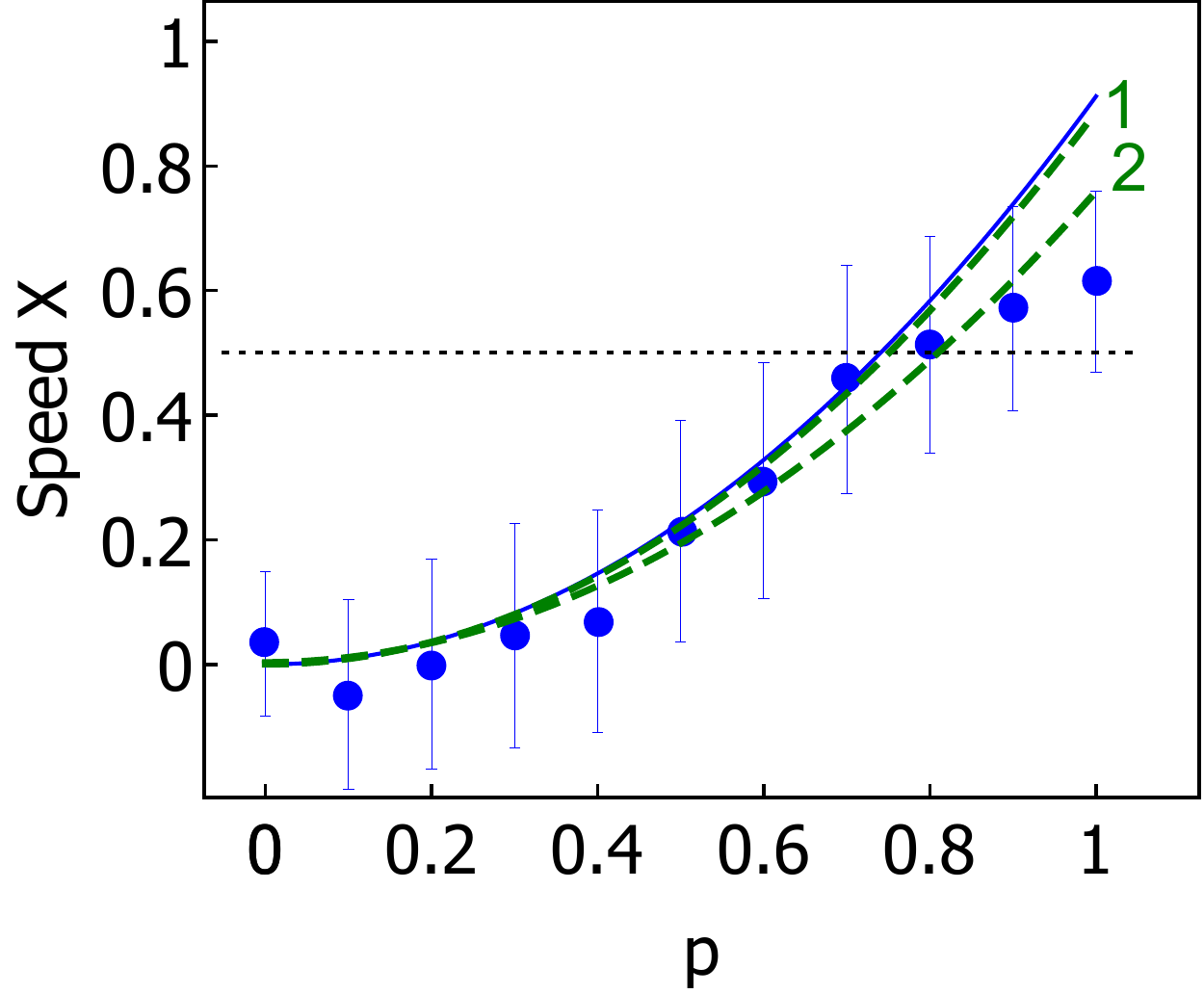}}\hspace{10pt}
 \subfigure[]{
\includegraphics[height=3.7cm,width=5.55cm]{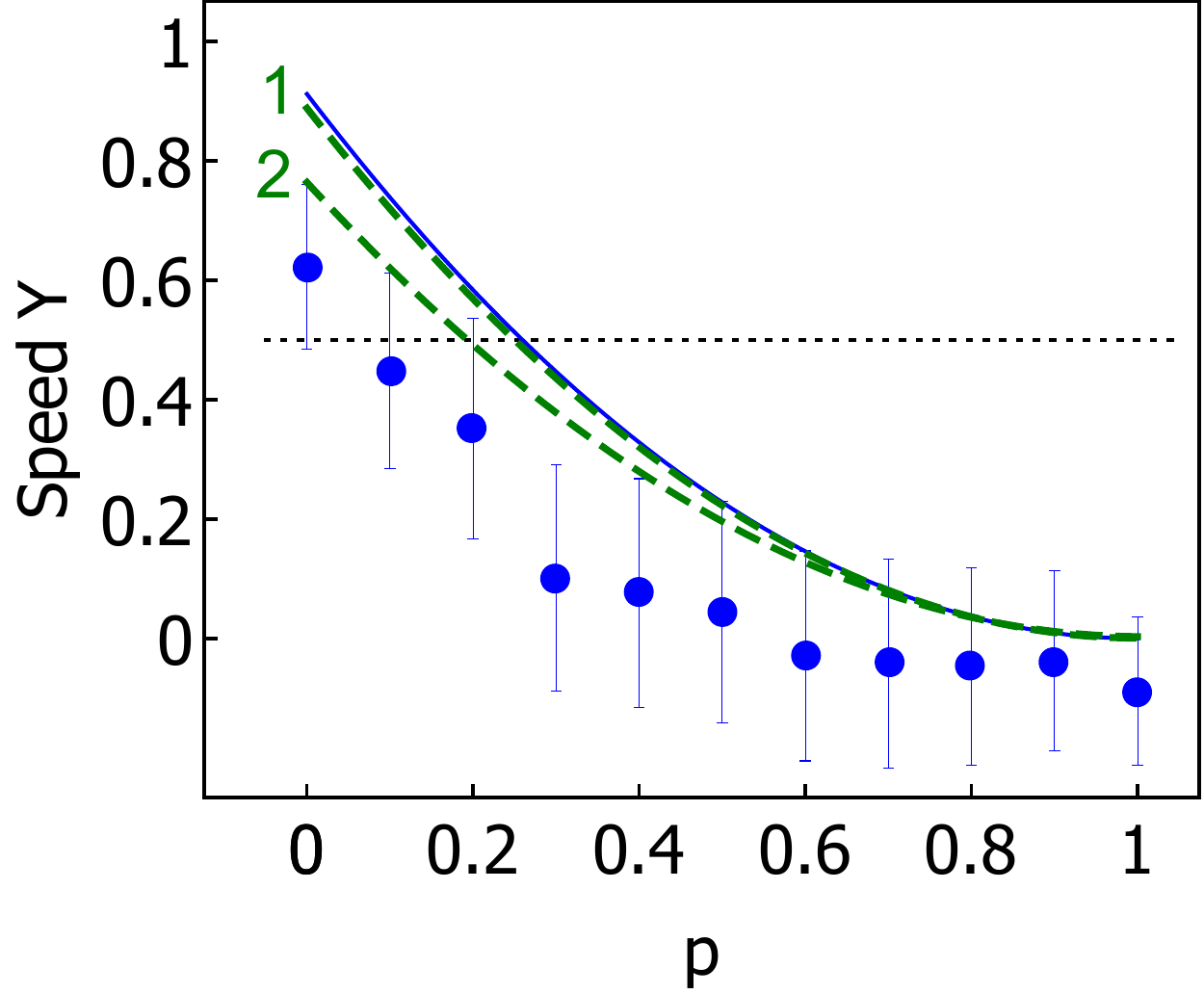}}\hspace{10pt}
 \subfigure[]{
\includegraphics[height=3.7cm,width=5.55cm]{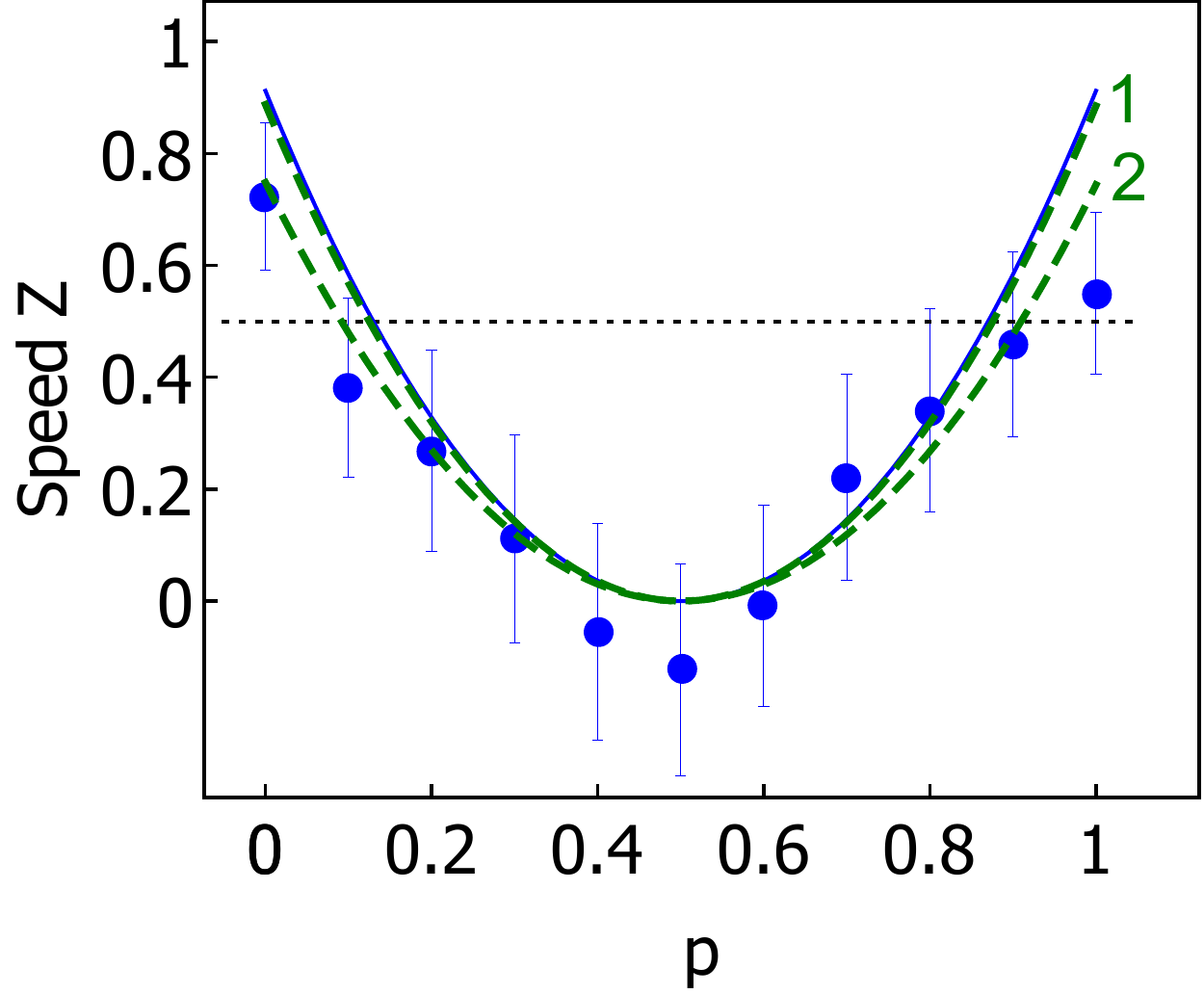}}
\caption{\label{fig3}{\bf Experimental results.} We evaluate the speed  of a two-qubit system in the state $\rho_p=p\ket{\phi_+}\bra{\phi_+}+(1-p)\ket{\phi_-}\bra{\phi_-},$ for unitary evolutions   $U_{\tau}\rho_p U^{\dagger}_{\tau}, U_{\tau}=e^{-i H_2\tau}, H_2=\sigma_{x,y,z \ A}\otimes I_B+I_A\otimes \sigma_{x,y,z \ B},$ over an interval $\tau=\pi/6$.  In Figs.~(a)-(c), the blue continuous line is the theoretical speed function ${\cal S}_{\pi/6}(\rho_p,H_2)$, which we aim at reconstructing, while the blue points are the  experimental values extracted from purity and overlap measurements, for $p=0,0.1,0.2,\ldots 0.9,1$. The error bars are determined by Monte Carlo simulation with Poisson-distributed error (1000 samples for each point). For comparison,  the two green dashed lines depict the speed function computed from the reconstructed states of Copy 1,2 (the density matrices are reported in the main text), respectively.   Super-linear scaling due to entanglement is detected for values above the horizontal, black dotted line.}
\end{figure*}

            The mean value of the swap operator is extracted by local and bi-local projections on the Bell singlet: $
 V_{12} =I_{12}-2\Pi^{\psi^-}_1\otimes I_2-2I_1\otimes \Pi^{\psi^-}_2+4 \Pi^{\psi^-\psi^-}_{12}$.
That is, three projections are required for evaluating purity and overlap, respectively. Note that for $n$ qubits $O(2^n)$ projections are required, still having exponential advantage with respect to full tomography.
 The projections are obtained via Bell state measurement (BSM) schemes applied to each subsystem pair. The BSMs consist of  PBSs,  HWPs, and photon detectors.  We place a $45^\circ$ HWP   in the input ports of the PBS corresponding to the $A_1,B_1$ subsystems to deterministically project into the Bell singlet \cite{BSM1}.  All the photons pass through single mode fibers for spatial mode selection. For spectral mode selection, photons 1-4 (5,6) pass through 3 nm (8 nm) bandwidth filters.\\
        The theoretical values to be extracted are given in Table~II.   The experimental results  are reported in Fig.~\ref{fig3}. For each Hamiltonian, we reconstruct the speed function ${\cal S}_{\pi/6}(\rho_p,H_2)$ from purity and overlap measurements, and compare it against the values obtained by  state tomography of the two system copies. By Eq.~(\ref{eq5}), entanglement is detected by  super-linear speed scaling ${\cal S}_{\pi/6}(\rho_p,H_2)\geq 1/2$. We observe that   speed values above the threshold detect entanglement yielding non-classical precision in phase estimation, not just non-separability of the density matrix (the state $\rho_p$ is entangled for $p\neq 1/2$).\\
        
  \subsection{Diagnostic of the experimental set-up}      
  \subsubsection{Error sources}
 We discuss the efficiency of the experimental set-up. The four photons interfering into the BSMs form a closed-loop network (Fig.~\ref{setup}).  This poses the problem to rule out the case of BSMs measuring two photon pairs emitted by a single SPDC source \cite{zeilinger}. We guarantee to generate the two system copies from different sources by preparing Copy 2 from two photon pair sources by post-selection. Single source double down conversion can also occur because of high order emission noise, which   has been minimised by setting a low  pump power.
   The coincidences have been  counted by a multichannel unit, with a 50/hour rate for about 6 hours in each experiment  run.  Here the main error source is the imperfection of the three Hang-Ou-Mandel  interferometers (one for the PBS and each BSM), which have a visibility of 0.91. This is due to the temporal distinguishability between the interfering photons, determined by the pulse duration.  The 3 nm and 8 nm narrow-band filters were placed in front of each detector to increase the photon overlap.

\subsubsection{Tomography of the input Bell state copies} 

We perform full state reconstruction of the two copies (Copy 1,2) of the  Bell states $\phi^{\pm}_{1,2}$ obtained by SPDC sources.    The fidelity of the input   states are respectively $0.9889 \ ({\phi}^+_{1})$, $0.9901 \ ({\phi}^-_{1})$, $0.9279 \ ({\phi}^+_{2})$, $0.9319 \ ({\phi}^-_{2})$.  We remind that Copy 1 (subsystems $A_1 B_1$) is generated by the sandwich-like Source 1 (photons 1,2), while Copy 2 ($A_2 B_2$) is  triggered by Sources 2,3 via parity check gate and post-selection applied to two product states (photons 3-6).  The counting rate for the Copy 1 photon pair is 32000/s, while for the four photons of Copy 2 is 110/s.   We use the maximum likelihood estimation method to reconstruct the related density matrices, which read:
  
\begin{eqnarray}
{\scriptsize
\hspace{-5pt}
  \begin{aligned}  
     \phi^+_{1}\!=\!\left(\!\!\!
          \begin{array}{cccc}
   0.5146 + 0.0000i  &-0.0158 + 0.0031i&   0.0058 + 0.0029i &  0.4923 + 0.0071i\\
  -0.0158 - 0.0031i &  0.0039 + 0.0000i & -0.0003 - 0.0026i  &-0.0173 - 0.0021i\\
   0.0058 - 0.0029i & -0.0003 + 0.0026i  & 0.0029 + 0.0000i  & 0.0029 - 0.0043i\\
   0.4923 - 0.0071i  &-0.0173 + 0.0021i &  0.0029 + 0.0043i &  0.4787 + 0.0000i\\
          \end{array}
        \!\!\!\right)\\
   \phi^-_{1}\!=
   \!\left(\!\!\!
          \begin{array}{cccc}
   0.5072 + 0.0000i  &-0.0065 + 0.0008i & -0.0052 + 0.0028i & -0.4931 - 0.0090i\\
  -0.0065 - 0.0008i &  0.0030 + 0.0000i &  0.0007 + 0.0021i  & 0.0065 + 0.0016i\\
  -0.0052 - 0.0028i &  0.0007 - 0.0021i &  0.0029 + 0.0000i &  0.0056 + 0.0034i\\
  -0.4931 + 0.0090i &  0.0065 - 0.0016i &  0.0056 - 0.0034i &  0.4869 + 0.0000i\\
          \end{array}
       \!\!\!\right)\\
      \phi^+_{2}\!=\!\left(\!\!\!
          \begin{array}{cccc}
   0.4881 + 0.0000i & -0.0108 + 0.0041i &  0.0063 + 0.0091i &  0.4486 + 0.0509i\\
  -0.0108 - 0.0041i  & 0.0216 + 0.0000i  &-0.0029 - 0.0066i & -0.0140 - 0.0068i\\
   0.0063 - 0.0091i  &-0.0029 + 0.0066i  & 0.0198 + 0.0000i &  0.0044 - 0.0073i\\
   0.4486 - 0.0509i  &-0.0140 + 0.0068i &   0.0044 + 0.0073i &  0.4706 + 0.0000i\\
          \end{array}
        \!\!\!\right)\\
    \phi^-_{2}
    \!=\!\left(\!\!\!
          \begin{array}{cccc}
   0.4911 + 0.0000i &  0.0041 - 0.0184i  & 0.0058 + 0.0075i & -0.4502 - 0.0462i\\
   0.0041 + 0.0184i &  0.0155 + 0.0000i  & 0.0005 + 0.0080i &  0.0041 - 0.0089i\\
   0.0058 - 0.0075i  & 0.0005 - 0.0080i  & 0.0209 + 0.0000i &  -0.0085 + 0.0182i\\
  -0.4502 + 0.0462i  & 0.0041 + 0.0089i  &-0.0085 - 0.0182i  & 0.4724 + 0.0000i\\
          \end{array}
       \!\!\!\right).\nonumber
  \end{aligned} }
\end{eqnarray}

\subsubsection{Tomography of  the Bell state measurements}
We analyse the efficiency of the measurement apparata.  A BSM consists of Hang-Ou-Mandel (HOM) interferometers and coincidence counts. The BSM is only partially deterministic, discriminating two of the four Bell states ($\ket{\phi^{\pm}}$, or $\ket{\psi^{\pm}}$) at a time.  The interferometry visibility in our setting is 0.91. Two BSM (1,2) are required to evaluate purity and overlap by measurements on two system copies. This requires the indistinguishability of the four interfering photons 1-4, including their arriving time, spatial mode and frequency.  As explained, our three source scheme ensures that, post-selecting sixfold coincidences,  each detected photon pair is emitted by a different source. We test our measurement hardware by performing BSM tomography. The probe states are chosen of the form $ \ket{\{{\bf H, V,D,A,R,L}\}}\bigotimes\ket{\{{\bf H, V,D,A,R,L}\}},$ where the labels identify the following photon polarisations:  horizontal ({\bf H}), vertical ({\bf V}), diagonal (${\bf D}=({\bf H+V})/\sqrt 2$), anti-diagonal $({\bf A}=({\bf H-V})/\sqrt 2)$,  right circular $({\bf R}= ({\bf H}+i {\bf V})/\sqrt 2)$, and left circular $({\bf L}=({\bf H}-i {\bf V})/\sqrt 2)$. The measurement results for all the possible outcomes are recorded accordingly.  An iterative maximum likelihood estimation algorithm yields the estimation of what projection is performed in each run \cite{MLE}.  
 The average fidelities of BSM1 and BSM2  are 0.9389 $\pm$  0.0030 and 0.9360 $\pm$ 0.0034, being the standard deviation calculated from  100 runs, by assuming Poisson statistics. The estimated Bell state projections $\Pi^{1(2)}_{x}=\ket{x}\bra{x}_{A_1(B_1)A_2(B_2)}, x=\phi^\pm,\psi^\pm,$ reconstructed from BSM1 (detecting on subsystems $A_1A_2$) and BSM2 (detecting on $B_1B_2$), are given by 
  
\begin{eqnarray}
{\scriptsize
\hspace{-5pt}
  \begin{aligned}
      \Pi_1^{\phi^+}\!&=\!\left(\!\!\!
          \begin{array}{cccc}
            0.5142& 0.0096 - 0.0102i & 0.0043 - 0.0055i & 0.4443 - 0.0088i \\
            0.0096 + 0.0102i &  0.0024  & -0.0005 + 0.0007i & -0.0037 + 0.0018i \\
            0.0043 + 0.0055i & -0.0005 - 0.0007i &  0.0052  &  0.0003 + 0.0110i \\
            0.4443 + 0.0088i & -0.0037 - 0.0018i &  0.0003 - 0.0110i &  0.4863  \\
          \end{array}
      \!\!\!\right)  \\ 
    \Pi_1^{\phi^-}
    \!&=\!\left(\!\!\!
          \begin{array}{cccc}
   0.4816  & -0.0088 + 0.0057i & -0.0081 + 0.0039i & -0.4481 + 0.0048i\\
  -0.0088 - 0.0057i &  0.0031  &  0.0013 + 0.0019i &  0.0136 - 0.0096i\\
  -0.0081 - 0.0039i &  0.0013 - 0.0019i &  0.0018  & -0.0001 - 0.0055i\\
  -0.4481 - 0.0048i &  0.0136 + 0.0096i & -0.0001 + 0.0055i &  0.5033 \\
          \end{array}
        \!\!\!\right)\\
      \Pi_1^{\psi^+}\!&=\!\left(\!\!\!
          \begin{array}{cccc}
   0.0014  & -0.0000 - 0.0083i &  0.0100 - 0.0010i &  0.0006 + 0.0006i\\
  -0.0000 + 0.0083i &  0.4954  &  0.4382 - 0.0059i & -0.0136 + 0.0147i\\
   0.0100 + 0.0010i &  0.4382 + 0.0059i &  0.5059  & -0.0057 + 0.0143i\\
   0.0006 - 0.0006i & -0.0136 - 0.0147i & -0.0057 - 0.0143i &  0.0014 \\
          \end{array}
        \!\!\!\right)\\
    \Pi_1^{\psi^-}
    \!&=\!\left(\!\!\!
          \begin{array}{cccc}
   0.0027  & -0.0008 + 0.0128i & -0.0062 + 0.0026i &  0.0032 + 0.0033i\\
  -0.0008 - 0.0128i &  0.4991  & -0.4390 + 0.0033i &  0.0038 - 0.0068i\\
  -0.0062 - 0.0026i & -0.4390 - 0.0033i &  0.4871  &  0.0054 - 0.0198i\\
   0.0032 - 0.0033i &  0.0038 + 0.0068i &  0.0054 + 0.0198i &  0.0090 \\
          \end{array}
       \!\!\! \right)\\
              \Pi_2^{\phi^+}\!&=\!\left(\!\!\!
          \begin{array}{cccc}
   0.4893  &  0.0043 - 0.0223i &  0.0064 - 0.0182i &  0.4397 - 0.0667i\\
   0.0043 + 0.0223i &  0.0017  &  0.0008 - 0.0004i &  0.0003 + 0.0159i\\
   0.0064 + 0.0182i &  0.0008 + 0.0004i &  0.0012  &  0.0123 + 0.0107i\\
   0.4397 + 0.0667i &  0.0003 - 0.0159i &  0.0123 - 0.0107i &  0.4942 \\
          \end{array}
       \!\!\!\right)\\
    \Pi_2^{\phi^-}
    \!&=\!\left(\!\!\!
          \begin{array}{cccc}
   0.5036  &  0.0050 - 0.0021i & -0.0015 + 0.0040i & -0.4413 + 0.0636i\\
   0.0050 + 0.0021i &  0.0023  & -0.0011 + 0.0008i &  0.0091 - 0.0072i\\
  -0.0015 - 0.0040i & -0.0011 - 0.0008i &  0.0011  & -0.0069 + 0.0007i\\
  -0.4413 - 0.0636i &  0.0091 + 0.0072i & -0.0069 - 0.0007i &  0.4987 \\
          \end{array}
        \!\!\!\right)\\
      \Pi_2^{\psi^+}\!&=\!\left(\!\!\!
          \begin{array}{cccc}
   0.0032  & -0.0098 + 0.0070i & -0.0140 + 0.0192i &  0.0018 + 0.0016i\\
  -0.0098 - 0.0070i &  0.4919  &  0.4375 + 0.0446i & -0.0101 - 0.0085i\\
  -0.0140 - 0.0192i &  0.4375 - 0.0446i &  0.5012  & -0.0059 - 0.0061i\\
   0.0018 - 0.0016i & -0.0101 + 0.0085i & -0.0059 + 0.0061i &  0.0050 \\
          \end{array}
        \!\!\!\right)\\
    \Pi_2^{\psi^-}
    \!&=\!\left(\!\!\!
          \begin{array}{cccc}
   0.0039  &  0.0005 + 0.0173i &  0.0091 - 0.0049i & -0.0001 + 0.0014i\\
   0.0005 - 0.0173i &  0.5041  & -0.4371 - 0.0451i &  0.0007 - 0.0002i\\
   0.0091 + 0.0049i & -0.4371 + 0.0451i &  0.4965  &  0.0004 - 0.0052i\\
  -0.0001 - 0.0014i &  0.0007 + 0.0002i &  0.0004 + 0.0052i &  0.0021 \\
          \end{array}
        \!\!\!\right).\nonumber
  \end{aligned} 
  }
\end{eqnarray}

\section{Theory background and full proofs}\label{five}
\subsection{Quantum Fisher informations as measures of state sensitivity}
Quantum Information geometry studies quantum states and channels as geometric objects. The Hilbert space of a finite $d$-dimensional quantum system admits a Rimenannian structure, thus it is possible to apply differential geometry concepts and tools to characterize quantum processes.  For an introduction to the subject, see Refs. \cite{apgeo,apamari}.\\
The information about a $d$-dimensional physical system is encoded in states represented by $d\times d$  complex hermitian matrices $\rho\geq0, \text{Tr}(\rho)=1,\rho=\rho^{\dagger},$ in the system Hilbert space ${\cal H}$. Each subset of rank $k$ states is a smooth manifold ${\cal M}^k({\cal H})$ of dimension $2 d k -k^2-1$ \cite{apmarmo}.  The set of all states ${\cal M}({\cal H})=\cup_{k=1}^d{\cal M}^k({\cal H})$  forms a stratified  manifold, where the stratification is induced by the rank $k$. The boundary of the manifold is given by the density matrices satisfying the condition $\det \rho=0$. \\
State transformations are represented on ${\cal M}({\cal H})$ as piecewise smooth curves $\rho: t\rightarrow \rho_{t}$, where $\rho_t$ represents the quantum state of the system at time $t\subseteq \mathbb{R}$.  By employing differential geometry techniques, it is possible to study  the space of quantum states ${\cal M}({\cal H})$ as a Riemannian structure.   The length of a  path $\rho_t, t \in[0,\tau],$ on the manifold is given by the integral of the line element
\begin{equation}
 l_{\rho_t}=\int_0^{\tau} d s=\int_{0}^{\tau} ||\partial_{t}\rho_{t}||\  dt, 
\end{equation}
 where the norm is induced by equipping ${\cal M}({\cal H})$ with a symmetric, semi-positive definite metric. The path length is invariant under monotone reparametrizations of the coordinate $t$.
 The definition of a metric function yields the notion of distance $d(\rho,\sigma)$ between two quantum states $\rho,\sigma$. The choice of the metric is arbitrary. However,  Morozova, Chentsov and Petz identified a class of functions, the quantum Fisher informations, which extend the contractivity of the classical Fisher-Rao metric under noisy operations  to quantum manifolds. This means that they have the appealing feature to be the unique class of  contractive Riemannian metrics under compeltely positive trace preserving (CPTP) maps $\Phi$: $d(\Phi(\rho),\Phi(\sigma))\leq d(\rho,\sigma), \forall \rho,\sigma,\Phi$ \cite{moro,geom}. For such a class of metrics, given the spectral decomposition of an input $\rho=\sum_i \lambda_i \ket{i}\bra{i}$, the line element associated to an infinitesimal  displacement $\rho\rightarrow \rho+d \rho$ takes the form
\begin{equation}
d s_{f}=\sqrt{\sum_i (d \lambda_i)^2 /4\lambda_i + \sum_{i< j} c_f(\lambda_i,\lambda_j)/ 2|\langle i|d\rho|j\rangle|^2}. 
\end{equation}
The terms $c_f(i,j)=(j f(i/j))^{-1}$, where the $f$s  are the Chentsov-Morozova functions \cite{moro}, identify the elements of the class.
We here describe their main properties, by focusing the analysis on the subclass of function identified by the regularity condition $f(0)>0$. The set of symmetric, normalised  Chentsov-Morozova operator monotones ${\cal F}_\text{op}$ consists of the real-valued functions $f: \mathbb{R}^+ \to \mathbb{R}^+$ such that
\begin{enumerate}
	\item[i)] For any hermitian operators $A,B$ such that $0 \leq A \leq B$, we have $0 \leq f(A) \leq f(B)$
	\item[ii)] $f(x) = x f(x^{-1})$
	\item[iii)] $f(1) = 1$.
\end{enumerate}
Thus, the following properties are satisfied:
\begin{enumerate}
\item[i)]$1/c_f(x,1):\mathbb{R}_+\rightarrow\mathbb{R}_+$
	\item[ii)]$c_f(x,y)=c_f(y,x),\  c_f(z x, z y)=z^{-1}c_f(x,y)$
	\item[iii)] $x \ c_f(x,1) = c_f(1/x,1)$
	\item[iv)]$x\leq y \Rightarrow  c_f(y,1)\leq c_f(x,1)$
	\item[v)]$c_f(1,1)=1$.
\end{enumerate}
By extending the domain of these functions to positive square matrices, they enjoy a one-to-one correspondence with the set ${\cal M}^m_\text{op}$ of \emph{matrix means} $m(A,B)$; see Ref.\ \cite{gibilisco2} for a list of defining properties. The link between the two sets is
\begin{equation}
	m_f(A,B) := A^{\frac{1}{2}} f(A^{-\frac{1}{2}} B A^{-\frac{1}{2}}) A^{\frac{1}{2}}, 
\end{equation}
which reduces to $m_f(A,B) = A f(BA^{-1})$ for commuting $A,B$.
Thus, matrix means also have a bijection with the set of  monotone Riemannian metrics which give rise to norms $|| A ||_{\rho,f}$ defined by
\begin{equation}
	|| A ||_{\rho,f}^2 := \text{Tr} \left( A\,m_f(L_\rho,R_\rho)^{-1}(A) \right),  
\end{equation}
where $R_\rho$ and $L_\rho$ are the right- and left-multiplication super-operators: $R_\rho(A) = A\rho,\, L_\rho(A) = \rho A$. The monotonicity property of these metrics implies contractivity under any CPTP map,
\begin{equation}
	||\Phi(A) ||_{\Phi(\rho),f} \leq ||A ||_{\rho, f}. 
\end{equation}
When applied to the stratified manifold of quantum states, such norms correspond to the quantum Fisher informations. Indeed,
 any metric defined on the manifold induces a metric on a parametrized curve $\rho_t=\sum_i \lambda_i(t)\ket{i(t)}\bra{i(t)}$. The squared rate of change at time $t$ is then given by the  tangent vector length
\begin{eqnarray} 
||\partial_t\rho_t||_{f}^2&=&\sum\limits_{i,j}\frac{|\langle i(t)|\partial_t\rho_t|j(t)\rangle|^2}{\lambda_j(t)f(\lambda_i(t)/\lambda_j(t))}\nonumber\\
&=& \sum_i (d_t \lambda_i(t))^2 /4\lambda_i(t)  \nonumber\\
&+& \sum_{i< j} c_f(\lambda_i(t),\lambda_j(t))/2|\langle i(t)|\partial_t\rho_t |j(t)\rangle|^2.  
\end{eqnarray}
The dynamics of the quantum Fisher informations for closed and open quantum systems has been studied in Ref. \cite{speedger}.\\
All such metrics reduce to the classical Fisher-Rao metric $\sum_i(d_{t}\lambda_i(t))^2/(\lambda_i(t))$ for stochastic dynamics of probability distributions $\{\lambda_i(t)\}$, represented at any time by a diagonal density matrix. On the other hand,  unitary transformations $\rho_{t}=U_{t}\rho U^{\dagger}_{t}$ are genuinely quantum,  as only the eigenbasis elements evolve. We focus on the latter case. Let us consider the unitary transformation $U_{t}\rho U^{\dagger}_{t}, U_{t}=e^{-i H t}$. The quantum Fisher  informations associated with $f \in {\cal F}_\text{op}$ read $\frac{f(0)}{2} ||i[\rho,H] ||_{f}^2$. We here absorb the constant factor and recast the quantity in the more compact form
\begin{equation}
 {\cal I}_f(\rho, H) :=1/4 ||i[\rho,H] ||_{f}^2. 
\end{equation}
For pure states, one has $2 f(0){\cal I}_f(|\psi\rangle\langle\psi|,H) = {\cal V}(|\psi\rangle\langle\psi|,H) = \langle H^2\rangle_{\psi}  - \langle H\rangle_{\psi}^2, \forall f$. For an arbitrary initial state  $\rho = \sum_i \lambda_i |i\rangle\langle i|$, it can be shown that
\begin{equation}
	{\cal I}_f(\rho,H) = \frac{1}{4} \sum_{i,j} \frac{(\lambda_i-\lambda_j)^2}{\lambda_j f(\lambda_i/\lambda_j)} |\langle i|H| j\rangle|^2, \end{equation}
where each term in the sum is taken to be zero whenever $\lambda_i = \lambda_j$  \cite{apgeo}.

\subsection{Proofs of theoretical results}
\subsubsection{Proof that any quantum Fisher information is an ensemble asymmetry monotone, extending the result in Eq.~(3)}
We prove two preliminary results upon which the result will be demonstrated.\\
\ \\
i) {\it  
	For any set of states $\rho_\mu$ and normalised probabilities $p_\mu$, and an orthonormal set $\{ |\mu\rangle \}$,}
\[
		{\cal I}_f\left(\sum_\mu p_\mu \rho_\mu \otimes |\mu\rangle\langle\mu|, H \otimes I\right) = \sum_\mu p_\mu {\cal I}_f(\rho_\mu, H), \forall f. 
\]
	 
		Let each $\rho_\mu$ have a spectral decomposition $\rho_\mu = \sum_i \lambda_{i|\mu}  |\psi_{\mu,i}\rangle\langle\psi_{\mu,i}|$. Recalling Eq.~(2), and defining $\lambda_{\mu,i} := p_\mu \lambda_{i|\mu}$, one has
		\begin{align*}
		&	{\cal I}_f\left(\sum_\mu p_\mu \rho_\mu \otimes  |\mu\rangle\langle\mu|, H \otimes I\right) =\nonumber\\
			&=\frac{1}{4} \sum_{\mu,\nu,i,j} \frac{(\lambda_{\mu,i}-\lambda_{\nu,j})^2}{\lambda_{\nu,j} f(\lambda_{\mu,i}/\lambda_{\nu,j})} |\langle\psi_{\mu,i}|\langle\mu| (H \otimes \mathbb{I}) |\psi_{\nu,j}\rangle |\nu\rangle|^2 \nonumber \\
			&=\frac{1}{4} \sum_{\mu,i,j} \frac{(\lambda_{\mu,i}-\lambda_{\mu,j})^2}{\lambda_{\mu,j} f(\lambda_{\mu,i}/\lambda_{\mu,j})} |\langle\psi_{\mu,i}|H|\psi_{\mu,j}\rangle|^2 \nonumber \\
			&=\frac{1}{4} \sum_{\mu,i,j} \frac{p_\mu^2 (\lambda_{i|\mu}-\lambda_{j|\mu})^2}{p_\mu \lambda_{j|\mu} f(\lambda_{i|\mu}/\lambda_{j|\mu})} |\langle\psi_{\mu,i}|H|\psi_{\mu,j}\rangle|^2 \nonumber \\
			&=\sum_\mu p_\mu {\cal I}_f(\rho_\mu,H). 
		\end{align*}
	
\noindent ii)	{\it ${\cal I}_f(\rho, H)$ is convex in $\rho$.} This follows from i), by tracing out the ancillary system, as ${\cal I}_f$ is monotonically decreasing under partial trace:
		\begin{eqnarray*}
			\sum_\mu p_\mu {\cal I}_f(\rho_\mu,H) &=& {\cal I}_f\left(\sum_\mu p_\mu \rho_\mu \otimes |\mu\rangle\langle\mu|\right)\\
			&\geq& {\cal I}_f\left(\sum_\mu p_\mu \rho_\mu,H\right). 
\end{eqnarray*}

We are now ready to prove determinsitic monotonicity. 
Recall that a $U(1)$-covariant channel, i.e. a symmetric operation, $\Phi$ is defined to be such that $[\Phi, U_t] = 0$, where $U_t(\rho) := e^{-i H t} \rho e^{i H t}$. Noting that $-i[H,\rho] = d_t U_t(\rho)|_{t=0}$, we have 
	${\cal I}_f(\rho,H) = \frac{f(0)}{2} || d_t U_t(\rho)|_0 ||_{f}^2.$ 
The linearity of $\Phi$ and the monotonicity property then give
\[
	|| d_t U_t(\Phi(\rho)) ||_{f} = || d_t \Phi(U_t(\rho)) ||_{f} = || \Phi(d_t U_t(\rho)) ||_{f}   \leq || d_t U_t(\rho) ||_{f},  
\]
so that ${\cal I}_f(\Phi(\rho),A) \leq {\cal I}_f(\rho,A), \forall f$.

To prove the ensemble monotonicity, we introduce a quantum instrument  as a set of covariant maps $\{\Phi_\mu\}$ which are not necessarily trace-preserving, while the sum $\sum_{\mu} \Phi_\mu$ is. For every quantum instrument, one can construct a trace-preserving operation by including in the output an ancilla that records which outcome was obtained via a set of orthonormal states $\{|\mu\rangle\}$, $
	\Phi'(\rho) := \sum_\mu \Phi_\mu(\rho) \otimes |\mu\rangle\langle\mu|$. 
Tracing out the ancilla results in the channel $\sum_\mu \Phi_\mu$. It is clear that $\Phi'$ is covariant whenever each of the $\Phi_\mu$ is. Writing $\Phi'(\rho) = \sum_\mu p_\mu \rho_\mu \otimes |\mu\rangle\langle\mu|$, result i) and deterministic monotonicity imply
\begin{align*}
	\sum_\mu p_\mu {\cal I}_f(\rho_\mu,H) & = {\cal I}_f\left(\sum_\mu p_\mu \rho_\mu \otimes |\mu\rangle\langle\mu|, H \otimes I\right) \nonumber \\
	& = {\cal I}_f(\Phi'(\rho), H \otimes I) \nonumber \\
	& \leq {\cal I}_f(\rho, H). 
\end{align*}

\subsubsection{Proof that the speed bounds any  quantum Fisher information, generalizing Eq.~(4)}
 
It is possible to express the system speed in terms of the Hilbert-Schmidt distance $D_{\text{HS}}(\rho,\sigma)=\sqrt{\text{Tr}((\rho-\sigma)^2)}$ and the related norm,
\[
{\cal S}_{\tau}(\rho,H)=D_{\text{HS}}^2(\rho, U_\tau \rho U_\tau^{\dagger})/(2\tau^2)=||U_\tau\rho U_\tau^{\dagger}-\rho||_{2}^2/(2\tau^2). 
\]
  The zero shift limit is given by
\[
 {\cal S}_{0}(\rho,H):=\lim\limits_{\tau->0}{\cal S}_\tau(\rho,H)=-1/2 \text{Tr}([\rho,H]^2). 
\]
By expanding the quantity in terms of the state spectrum and eigenbasis, one has ${\cal S}_0(\rho,H)= \sum_{i\neq j}(\lambda_i -\lambda_j)^2/2 |\langle i|H|j\rangle|^2$. \\
We recall the norm inequality  chain
$ f(0)/2 \ ||A||_f\leq1/4||A||_F\leq1/4||A||_f,\ \forall f, A, $ which, for unitary transformations $e^{-i H t}\rho e^{i H t}$, implies the topological equivalence of the quantum Fisher informations:
\[
2f(0) {\cal I}_f(\rho,H)\leq{\cal I}_F(\rho,H)\leq  {\cal I}_f(\rho,H),\ \forall f, \rho,H, 
\]
where $F$ labels the  SLDF \cite{gibilisco2}.  We note that its expansion for unitary transformations reads ${\cal I}_F(\rho,H)=\sum_{i\neq j}(\lambda_i -\lambda_j)^2/(2(\lambda_i+\lambda_j)) |\langle i|H|j\rangle|^2$. Since   $\lambda_i+\lambda_j \leq 1, \forall i,j $, it follows that
\[
{\cal S}_0(\rho,H)\leq {\cal I}_F(\rho,H), \forall \rho,H. 
\]
Any distance between two states is defined as the length of the shortest path between them. By recalling the von Neumann equation $\partial_t\rho_t=i[\rho_t,H]$, and integrating over the unitary evolution $U_{t}$, one  obtains
\begin{align*}
D_{\text{HS}}(\rho, U_\tau \rho U_\tau^{\dagger})
&\leq\int_{\rho_t\equiv\rho}^{\rho_t\equiv U_\tau \rho U_\tau^{\dagger}}||\partial_t \rho_t||_{2}\ d t\nonumber\\
&\leq\int_{\rho_t\equiv\rho}^{\rho_t\equiv U_\tau \rho U_\tau^{\dagger}}\left(-\text{Tr}([\rho,H]^2)\right)^{1/2}d{t}\nonumber\\
&=\int_{\rho_t\equiv\rho}^{\rho_t\equiv U_\tau \rho U_\tau^{\dagger}}\left(2{\cal S}_0(\rho,H) \right)^{1/2}d{t}\nonumber\\
&= \left(2{\cal S}_0(\rho,H) \right)^{1/2} \tau\nonumber\\
&\leq \left(2{\cal I}_F(\rho,H) \right)^{1/2}\tau\nonumber\\
&\leq \left(2{\cal I}_f(\rho,H) \right)^{1/2}\tau,\forall f.  
\end{align*}
 Hence, the bound is proven. The inequality is saturated for pure states in the limit $\tau\rightarrow 0$.

\subsubsection{Bonus: Determining the scaling of the SLDF from speed measurements for pure states mixed with white noise}
Suppose we are given the state
\begin{equation*}
	\rho_{\epsilon} = (1 - \epsilon) \rho_{\psi} + \epsilon \frac{I_d}{d},
\end{equation*}
where $I_d$ is the identity of dimension $d$, while $\rho_{\psi}$ is an arbitrary pure state and $\epsilon$ is unknown.  
By convexity, one has
\begin{equation*}
	{\cal I}_F(\rho_\epsilon,H) \leq (1-\epsilon){\cal I}_F(\kebra{\psi}{\psi},H)\leq (1-\epsilon){\cal S}_0(\kebra{\psi}{\psi},H), \forall H, 
\end{equation*}
since $I_d/d$ is an incoherent state in any basis. By Eq.(2), few algebra steps give

\begin{equation*}
	{\cal S}_0(\rho_\epsilon,H) \leq {\cal I}_F(\rho_\epsilon,H) \leq \sqrt{\frac{d-1}{d \text{Tr}(\rho_\epsilon^2)- 1}} {\cal S}_0(\rho_\epsilon,H). 
\end{equation*}
By Taylor expansion about $\tau =0$, one has ${\cal S}_\tau(\rho,H)={\cal S}_0(\rho,H)+O(\tau^2), \forall \rho, H$. Thus, measuring the speed function  ${\cal S}_{\tau}(\rho_\epsilon)$ and the state purity determines both upper and lower bounds to the SDLF, and consequently to any quantum Fisher information, up to an experimentally controllable error due to the selected time shift. 

\section{Conclusion}\label{six}
  We showed how to extract quantitative bounds to metrologically useful asymmetry and entanglement in multipartite systems from a limited number of measurements, demonstrating the method in an all-optical experiment.  The scalability of the  scheme may make possible to certify quantum speed-up in large scale registers  \cite{toth,nielsen,entropy}, and to study  critical properties of many-body systems \cite{zanardi,critic2,zoller}, by limited laboratory resources.  On this hand, we remark that we here compared our method with state tomography, as the two approaches share the common assumption that no a priori knowledge about the input state and the Hamiltonian is given. An interesting  follow-up work would  test the efficiency of our entanglement witness against two-time measurements of the classical Fisher information, when local measurements on the subsystems are only available.  A further  development would be to investigate  macroscopic quantum effects via  speed detection, as they have been  linked to quadratic precision scaling in phase estimation (${\cal I}_F(\rho,H_n)=O(n^2)$)  \cite{jeong,frowis,ben}.

\section*{Acknowledgements}
 We thank Paolo Gibilisco, Tristan Farrow, and an anonymous Referee for fruitful comments. This work was supported by: the Oxford Martin School and the Wolfson College, University of Oxford;  the EPSRC (UK) Grant EP/L01405X/1; the Leverhulme Trust (UK); the John Templeton Foundation; the EU Collaborative Project TherMiQ (Grant Agreement 618074); the COST Action MP1209;  the NRF, Prime Ministers Office, Singapore; the Ministry of Manpower (Singapore) under its Competitive Research Programme (CRP Award No. NRF- CRP14-2014-02), administered by the Centre for Quantum Technologies (NUS); the  Key Research Program of Frontier Sciences, CAS (No. QYZDY-SSW-SLH003); the National Natural Science Foundation of China (Grant Nos. 11274289, 11325419, 11374288, 11474268, 61327901, 61225025 and 61490711); the Strategic Priority Research Program (B) of the Chinese Academy of Sciences (Grant No. XDB01030300); the Fundamental Research Funds for the Central Universities, China (Grant Nos. WK2470000018 and WK2470000022); the National Youth Top Talent Support Program of National High-level Personnel of Special Support Program (No. BB2470000005).     
    
\end{document}